# Scaling Nanoribbon Transistors with Monolayer Transition Metal Dichalcogenides

Tara Peña[1†*], Anton E. O. Persson[1,2†*], Andrey Krayev[3], Áshildur Friðriksdóttir[4], Haotian Su[1,5], Yuan-Mau Lee[4], Young Suh Song[1], Kathryn Neilson[1], Zhepeng Zhang[4], Anh Tuan Hoang[1,4], Jerry A. Yang[1], Lauren Hoang[1], Shan X. Wang[1,4,5], Andrew J. Mannix[4,6], Paul C. McIntyre[4,7,8], and Eric Pop[1,4,7,8,9,*]

[1]*Department of Electrical Engineering, Stanford University, Stanford, CA 94305, USA*
[2]*Department of Microtechnology and Nanoscience, Chalmers University of Technology, 412 96 Gothenburg, Sweden*
[3]*HORIBA Scientific, Novato, CA 94949, USA*
[4]*Department of Materials Science & Engineering, Stanford University, Stanford, CA 94305, USA*
[5]*Geballe Laboratory for Advanced Materials, Stanford University, Stanford, CA 94305, USA*
[6]*Stanford Institute for Materials and Energy Sciences, SLAC National Accelerator Laboratory, Menlo Park, CA 94025, USA*
[7]*Stanford Synchrotron Radiation Lightsource (SSRL), SLAC National Accelerator Laboratory, Menlo Park, CA 94025, USA*
[8]*Precourt Institute for Energy, Stanford University, Stanford, CA 94305, USA*
[9]*Department of Applied Physics, Stanford University, Stanford, CA 94305, USA*

[*]*Corresponding authors: Tara Peña, tara.pena@stanford.edu; Anton E. O. Persson, anton.persson@chalmers.se; Eric Pop, epop@stanford.edu*
[†]*Authors contributed equally to this work.*

Nanoscale transistors demand aggressive scaling of all channel dimensions—length, width and thickness. Two-dimensional semiconductors (2DS) provide the ultimate thickness limit, yet good device performance has largely remained restricted to micrometre-wide channels. Here we report monolayer-2DS nanoribbon transistors with both *n*- and *p*-type operation, fabricated by a top-down multi-patterning process that includes 'anchored' contact pads to limit nanoribbon delamination. This approach achieves channel lengths and widths down to 25-30 nm, with minimal edge degradation confirmed through nanoscale characterization, including tip-enhanced photoluminescence. Integrated with thin high-κ gate dielectrics, the devices deliver on-state currents up to 560, 420 and 130 μA μm$^{-1}$ at 1 V drain-to-source voltage for *n*-type $MoS_2$, $WS_2$ and *p*-type $WSe_2$, respectively. These results exceed prior single-gated 2DS nanoribbon reports, with $WS_2$ improving by more than two orders of magnitude, even for normally off (enhancement-mode) operation. Overall, these findings position top-down patterned 2DS nanoribbons as promising building blocks for future nanosheet transistor architectures.

## Introduction

The history of transistors for digital computing has experienced only three major changes in device architecture: the transition from bipolar to metal-oxide semiconductor field-effect transistors (FETs)[1] in the 1970s, the transition to FinFET or tri-gate transistors[2] around 2007, and the present transition to gate-all-around (GAA) nanosheets[3] in 2025. While silicon-based GAA transistors are expected to scale for at least another decade, it is unclear if the further thickness reduction required below 3 nm to retain electrostatic control is feasible due to degradation of electrical properties[4–6]. Instead, two-dimensional (2D) semiconductors, like monolayer transition metal dichalcogenides (TMDs), are appealing alternatives due to their good electrical properties (*e.g.,* mobility, band gap) in sub-nanometer thin films[7,8] and their potential in scaled GAA devices[9]. Accordingly, 2D transistors have been recently placed on technology roadmaps[10], with targeted integration by the late 2030s or early 2040s.

The most important building block of GAA nanosheet transistors is the nanoribbon channel, which must be between 10 to 50 nm wide[11] and atomically thin for the best electrostatic gate control[12,13]. Monolayer 2D semiconductors with sub-nanometer thickness should also enable shorter, sub-5 nm, gates[14,15] than thicker silicon nanosheets[7], making them promising for continued scaling and higher device density. However, to date, most demonstrations of good performance in monolayer 2D TMD transistors have used short, sub-100 nm, but micrometer-wide channels. The reasons are manifold but likely include difficulty in fabrication (*e.g.,* TMD delamination, lithography limitations), difficulty in making good contacts, and mobility degradation due to edge imperfections. Little is known, for example, about how or if charge transport in narrow TMD ribbons changes in channel widths below a micrometer, and there are concerns about magnified edge effects in such devices[16,17].

Here we tackle the challenges mentioned above (adhesion, width scaling, contacts, and edge roughness) by realizing *n*- and *p*-type monolayer TMD nanoribbons with similar performance as co-fabricated micrometer-wide control devices. A key advance is anchoring the contacts to the substrate during fabrication, which limits nanoribbon delamination and cracking, allowing us to study many such devices. We also introduce a multi-patterning approach to achieve nanoribbon widths down to 25 nm. With these advances, we reach high current density in monolayer $MoS_2$ nanoribbons, over 600 μA $\mu m^{-1}$ with $SiO_2$ gate dielectric (560 μA $\mu m^{-1}$ with $HfO_2$ dielectric) at 1 V drain-to-source bias. We also achieve the highest saturation current density in monolayer $WS_2$ nanoribbons to date, over 450 μA $\mu m^{-1}$ in enhancement mode, normally off devices. Imaging the nanoribbons and their edges by tip-enhanced photoluminescence (TEPL) and transmission electron microscopy (TEM) suggests that edge disorder is not the limiting factor at these dimensions, indicating that such top-down monolayer TMDs are promising candidates for future nanosheet transistors.

## Fabrication of anchored nanoribbons

Due to lack of out-of-plane chemical bonds, monolayer TMDs adhere to substrates by van der Waals forces, making them prone to delamination during lithography, etching, and wet processing. To mitigate this, we designed a dog-bone-shaped structure where the TMD is narrow only in the channel but expands into wider pads under the source and drain contacts (**Fig. 1a**). These micrometer-sized regions anchor the nanoribbon to the substrate, increasing the mechanical stability and the reproducibility when reducing the nanoribbon widths. The approach is somewhat analogous to the wider source and drain regions used for silicon nanosheets to enable channel release and reduce contact resistance[18].

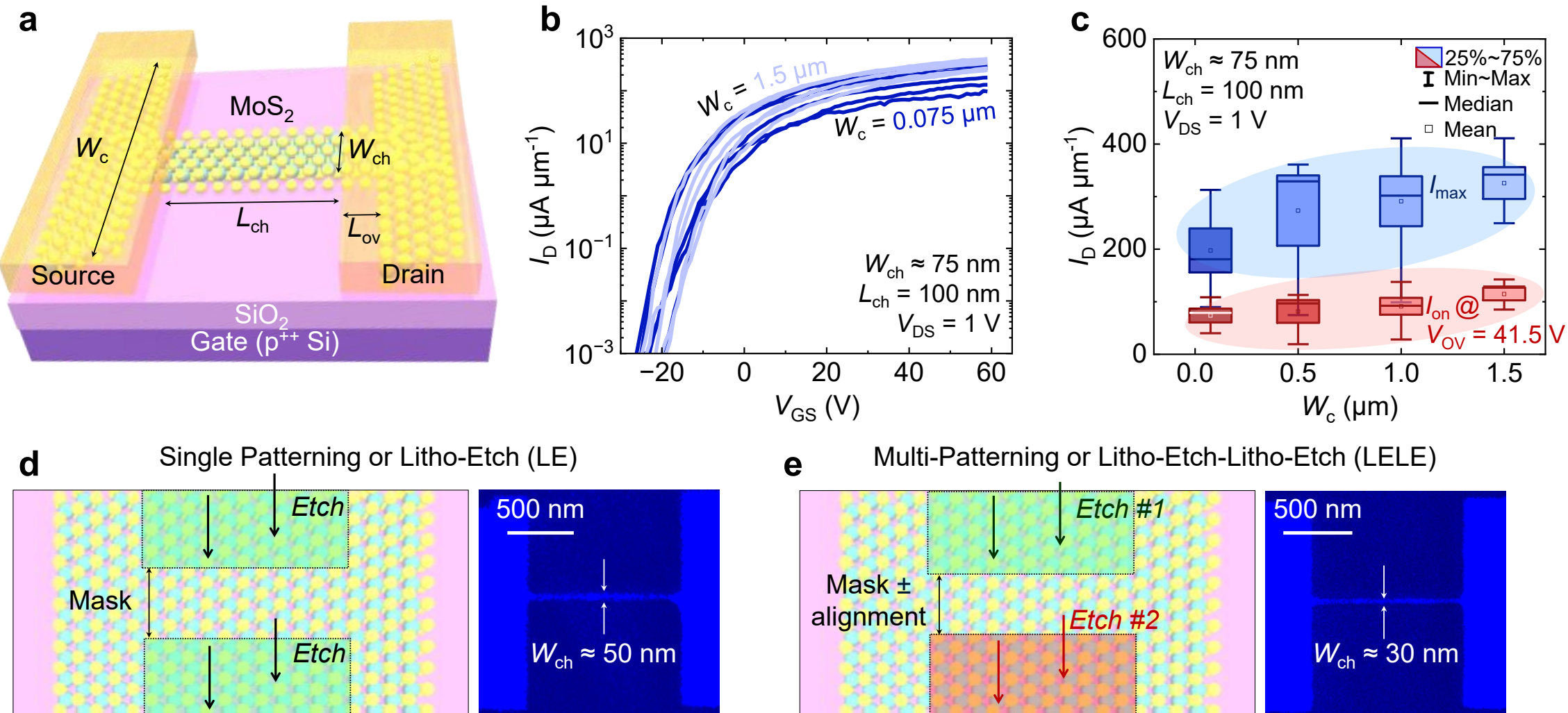

**Figure 1 | Anchored contacts and multi-patterning for improved device fabrication. a**, Schematic of the dog-bone shaped back-gated monolayer $MoS_2$ transistor, defining the channel width ($W_{ch}$), length ($L_{ch}$), contact width ($W_c$), and contact overlap region ($L_{ov}$). **b**, Transfer characteristics of various short-channel devices (100 nm), displaying larger contact widths in light blue (1.5 µm) and smaller contact widths in dark blue (75 nm). **c**, Box plots of maximum drain current density ($I_{max}$) and the drain current at fixed gate overdrive ($I_{on}$ at $V_{ov} = V_{GS} - V_T$) for varying contact widths. Each box plot includes five to six devices, totaling 22 tested devices. **d**, Schematic of lithography-etch (LE) single-patterning approach (left), which can define nanoribbon widths down to ~50 nm with reduced electron-beam dose. Representative false-colored scanning electron microscopy (SEM) image (right) of a resulting nanoribbon. **e,** Schematic of lithography-etch-lithography-etch (LELE) multi-patterning strategy (left, also see **Extended Data Fig. 1** for more details), used to achieve nanoribbon widths below 50 nm. False-colored SEM image (right) shows a nanoribbon with ~30 nm width.

We first study $MoS_2$ nanoribbons on conventional $SiO_2$ (96 nm) on $p^{++}$ Si substrates, which also serve as back-gates. This allows more rapid process optimization, because the monolayer $MoS_2$ is grown directly on the $SiO_2$, enabling higher throughput and better adhesion than layer-transferred films.

As shown in **Fig. 1b**, short-channel devices with $W_{ch} \approx 75$ nm width and $W_c$ of either 75 nm or 1.5 μm display nearly identical transfer curves. Encouragingly, we observe good maximum current, up to $I_{max} \approx 400$ μA μm$^{-1}$ at $V_{DS} = 1$ V, and off-state currents limited by the measurement noise floor. **Figure 1c** shows the current at fixed gate overdrive, $I_{on}$ at $V_{ov} = V_{GS} - V_T$, remains nearly unchanged while $W_c$ increases by a factor of 20. ($V_T$ is the threshold voltage, see **Supplementary Fig. 1b**.) We attribute this behavior to two factors: (i) the contact-channel overlap ($L_{ov} > 100$ nm) exceeds the expected current[19] and thermal transfer length of the contacts, and (ii) a fabrication process that minimizes patterning damage at sub-100 nm widths, which we discuss below. (We estimate the temperature profile along the nanoribbons in **Supplementary Fig. 2**.) Thus, the dog-bone structure preserves the nanoribbon transistor behavior, while greatly improving our yield due to contact anchoring. Without it, sub-100 nm wide nanoribbons often delaminate during processing, while the anchored contacts enabled >85% yield down to 60 nm widths (**Supplementary Fig. 1c**). For these reasons, the nanoribbons investigated in the remainder of the manuscript use the wider $W_c = 1.5$ μm, although we note that industrial manufacturing will require alternative yield improvement methods compatible with smaller contact areas[18,20].

Although single-step lithography-and-etching (LE, **Fig. 1d**) is commonly used to pattern the 2D channel in academic studies, we wanted to limit the electron-beam exposure dose to reduce the density of lithographically-induced defects[21,22]; thus, this method reaches a limit of ~50 nm widths due to the reduced dose and other fabrication trade-offs (see **Methods**). To make narrower ribbons, we employ a litho-etch-litho-etch (LELE) multi-patterning approach inspired by modern industrial lithography (**Fig. 1e** and **Extended Data Fig. 1**). This enables nanoribbons down to ~25 nm widths while maintaining the same overall low dose as the single-patterning LE approach (see **Methods**). Achieving sub-25 nm wide nanoribbons should be possible by optimizing the TMD adhesion and anchoring, however, sub-15 nm nanoribbons may not be desirable for GAA transistors due to larger parasitic capacitance[4,23].

**Electrical characterization of $MoS_2$ nanoribbon transistors**

We also investigate the effect of contact resistance, by evaluating nanoribbon behavior as a function of channel length using ten transfer-length-method (TLM) structures. These nanoribbons have $W_{ch} \approx 75$ nm and 'anchored' contacts with $W_c \approx 1.5$ μm. As shown in **Fig. 2a**, the devices exhibit excellent channel length dependence, maintaining stable characteristics and achieving on-state current density up to ~400 μA μm$^{-1}$ at channel lengths $L_{ch} \approx 300$ nm and $V_{DS} = 1$ V. **Figure 2b** displays the total device resistance $R_{tot}$ vs. channel length and the linear extrapolation yields a contact resistance $R_c < 560$ Ω·μm, or 190 ± 370 Ω·μm from the linear fit at the highest $V_{ov}$, comparable to the best $MoS_2$/Au contacts reported to date[6]. We note that the $R_c$ and $R_{tot}$ are normalized by the channel width $W_{ch}$, not by $W_c$, because the contacts have a non-negligible overlap with the channel ($L_{ov}$ > 100 nm, greater than the expected current transfer length at these contacts[6,19]), as shown in **Fig. 1a**.

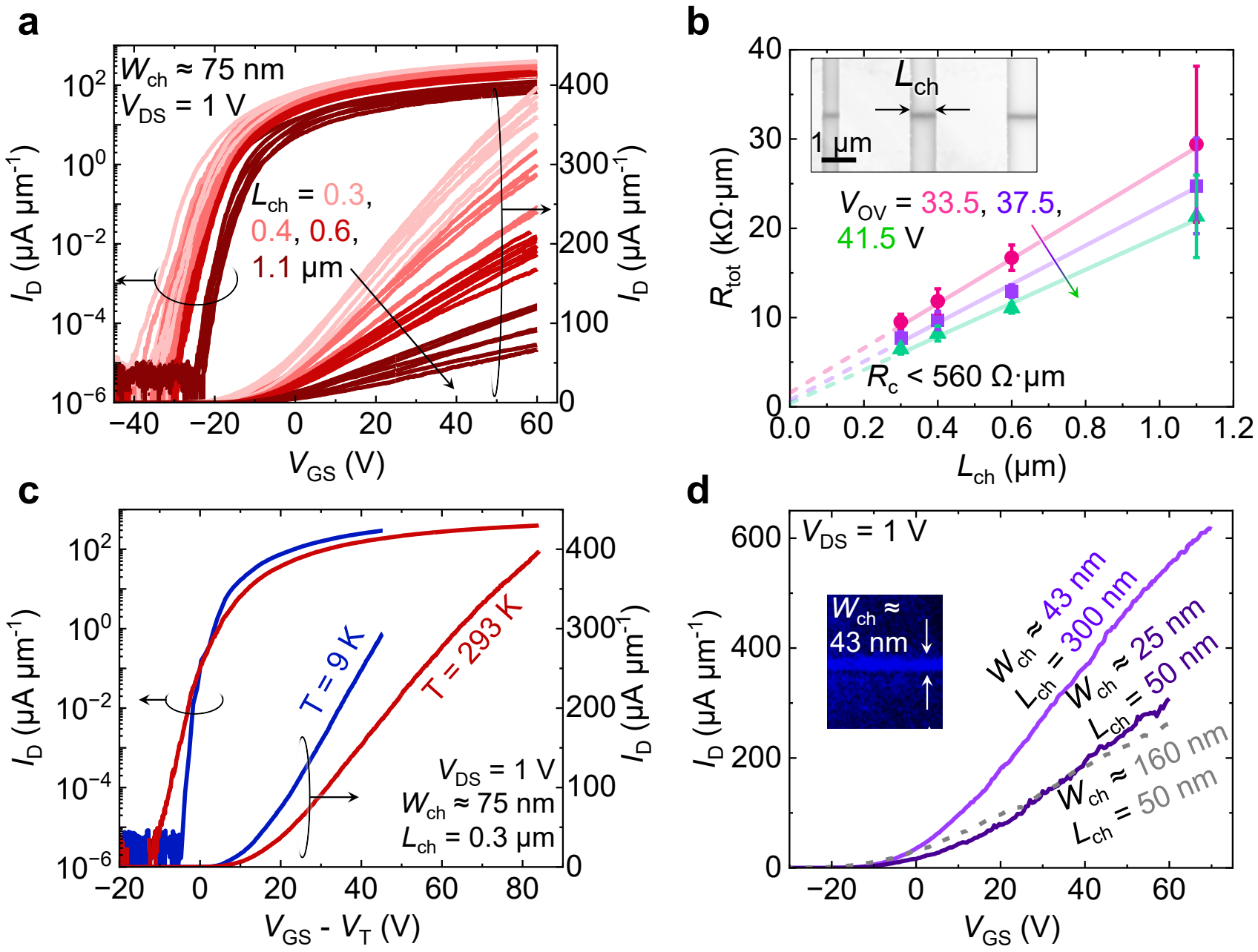

**Figure 2 | Monolayer $MoS_2$ nanoribbons on $SiO_2$. a**, Transfer characteristics of nanoribbons with four different channel lengths, showing high on-currents and consistent device behavior. Between 6 and 10 nanoribbons are measured for each length, 30 devices total. $W_{ch}$ = 75 nm and $W_c$ = 1.5 μm here. **b**, Transfer-length-method (TLM) analysis of the same devices; each symbol and error bar represent an average and standard deviation of devices with the same channel length. The contact resistance is comparable to state-of-the-art $MoS_2$/Au contacts reported[6] in literature ($R_c$ < 560 Ω·μm at the highest $V_{ov}$). Inset: scanning electron microscopy (SEM) image of such a TLM structure. **c**, Transfer characteristics of a nanoribbon device at low temperature (9 K) compared to room temperature, as a function of $V_{GS} - V_T$ overdrive; $V_T$ is taken here at a constant current of 100 nA μm$^{-1}$. **d**, Transfer characteristics of other nanoribbons, at $V_{DS}$ = 1 V. The two with 43 nm (LE) and 25 nm (LELE) channel width reach $I_{max}$ ≈ 620 μA μm$^{-1}$ and ~310 μA μm$^{-1}$, respectively, the highest current density reported to date for single-gated monolayer TMD nanoribbons at these widths. The 160 nm wide nanoribbon (dashed grey line, also on LELE chip) has similar current density as the 25 nm wide device, suggesting these were limited by higher contact resistance rather than edge disorder. The inset is a false-colored SEM of the 43 nm wide nanoribbon device.

Interestingly, we do not see mobility degradation in nanoribbons (~75 nm wide) compared to much wider devices (~850 nm wide) fabricated on the same chip. As summarized in **Supplementary Fig. 3**, the field-effect electron mobility $\mu_{FE}$ is in the range of 30 to 60 cm$^2$ V$^{-1}$ s$^{-1}$ for ten devices at room temperature, with greater device-to-device variation than any apparent width-dependence. This is not surprising, because the intrinsic electron mean free path in monolayer $MoS_2$ is expected[24] to be 3 to 5 nm, much shorter than the nanoribbon width. Nevertheless, it is reassuring that the edges produced from the top-down patterning process used here does not appear to deteriorate device operation, a topic we return to below. **Figure 2c** displays low-temperature measurements of a nanoribbon at ~9 K ambient, revealing that mobility approximately doubles, which suggests that low-temperature transport is ultimately limited by impurities and possibly by the edges (see **Supplementary Fig. 4**).

We also show transfer characteristics of other nanoribbons in **Fig. 2d**; one has $W_{ch} \approx 43$ nm and $L_{ch} \approx 300$ nm, using the single-step LE approach, and two others have $W_{ch} \approx 25$ nm and 160 nm with $L_{ch} \approx 50$ nm, using the LELE multi-patterning approach. The first two reach $I_{max} \approx 620$ μA μm$^{-1}$ and ~310 μA μm$^{-1}$, respectively, at $V_{DS} = 1$ V and similar $V_{GS}$; the former is the highest current density to date in single-gated $MoS_2$ nanoribbons; the latter is the highest for any monolayer ribbon of such small width. The 25 nm and 160 nm wide nanoribbons on the same LELE chip have similar current density, suggesting they are not limited by edge disorder. However, their $I_{max}$ is lower in a shorter channel than the device on the LE chip, which we attribute to higher contact resistance arising from our chip-to-chip variation (also see **Supplementary Fig. 5**). The nanoribbon width estimates have a 3 to 5 nm uncertainty (see **Methods** details), which implies 10 to 20% uncertainty of current density.

**Nanoscale spectroscopic and structural characterization**

We next wish to understand the origin of good performance in our nanoribbon devices and to evaluate the fabrication process and the effect of the edges. Raman spectra shown in **Fig. 3a** for representative monolayer $MoS_2$ devices reveal no discernible signs of damage, as the predominant E' (in-plane) and $A'_1$ (out-of-plane) phonon modes do not broaden with reduced nanoribbon widths, down to 45 nm. We also do not observe defect-mediated phonon modes[25,26] [e.g., LA(M) at 227 cm$^{-1}$] in any of our nanoribbons, suggesting that edge-related defects are not predominant. We also map our nanoribbons with tip-enhanced photoluminescence (TEPL), as shown in **Fig. 3b,c**, which reveal uniform TEPL across a long and narrow channel ($L_{ch} \approx 1$ μm, $W_{ch} \approx 75$ nm). The TEPL spectra confirm the quality of the nanoribbons, wherein the A-exciton does not broaden compared to the wider regions. The nanoribbons do exhibit a slightly larger trion-to-exciton ratio (**Supplementary Fig. 6**), which potentially suggests doping from the edges (see **Extended Data Fig. 4** for more discussion).

To visualize the atomic structure and edge roughness, we performed high-angle annular dark-field scanning transmission electron microscopy (HAADF-STEM) and monochromated transmission electron microscopy (TEM) imaging on monolayer $MoS_2$ nanoribbons transferred onto 10 nm thick $SiN_x$ membranes (**Fig. 3d,e**). Compared to the nanoribbon transistors, these samples may be subject to some additional damage from the transfer process and the electron beam exposure during imaging. Despite this, the nanoribbon edges appear clean, with edge roughness of at most a few nanometers. The edge termination likely alternates between zigzag and armchair segments, as expected from a top-down patterning process without edge-selective anisotropy. Energy-dispersive X-ray spectroscopy (EDS) and electron energy loss spectroscopy (EELS) reveal no discernible accumulation of oxygen or fluorine near the edges in monolayer $MoS_2$ and $WSe_2$ (**Supplementary Figs. 7-9**). These findings suggest that our top-down fabrication approach does not introduce observable disorder or contamination of the nanoribbons and their edges, supporting the lack of degradation seen in electrical characteristics.

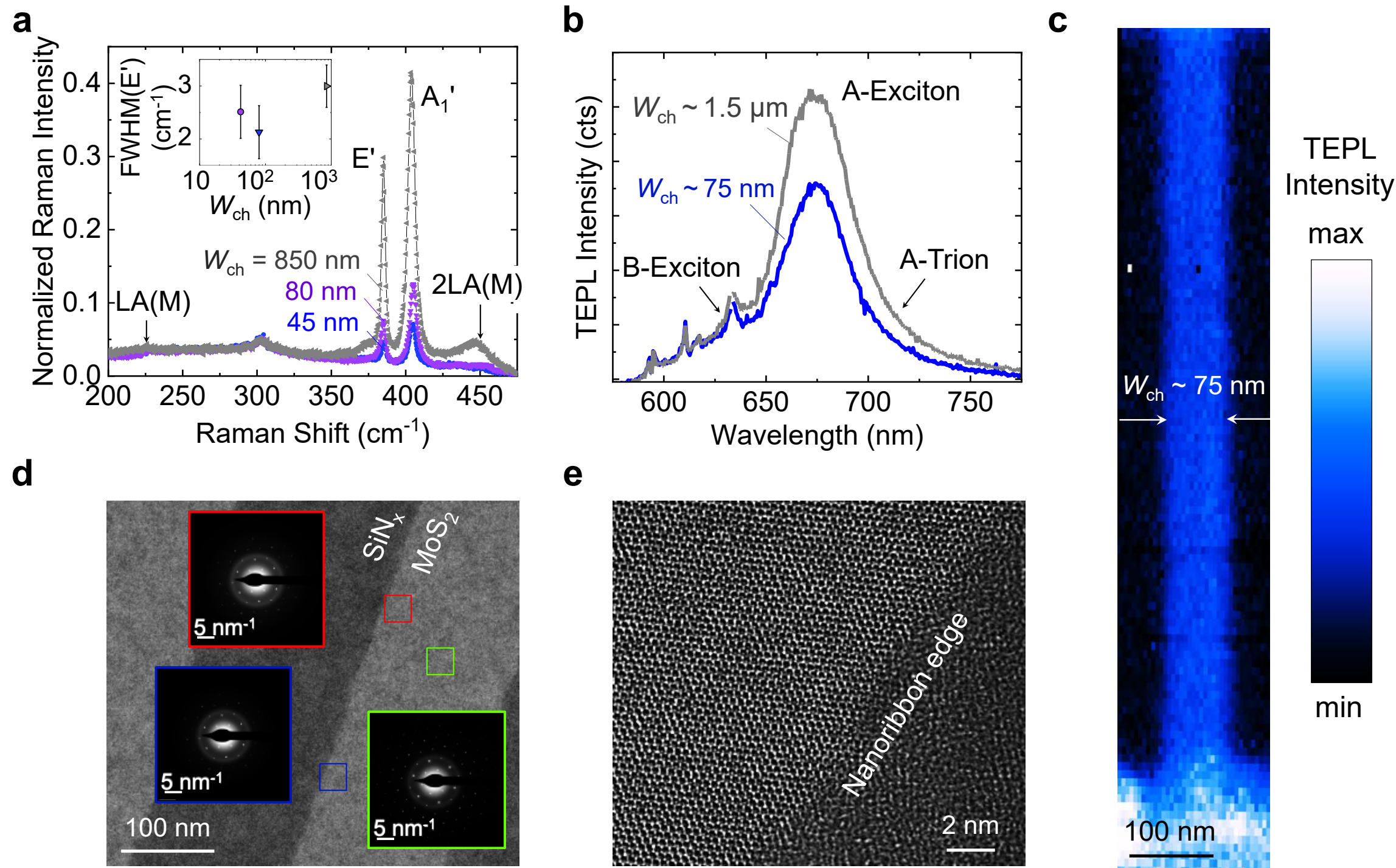

**Figure 3 | Nanoribbon material and edge characterization. a**, Raman spectra of monolayer $MoS_2$ device channels between ~45 nm to 850 nm wide, showing no discernible broadening of the E' or $A'_1$ modes, nor contributions from LA(M) defect-mediated peak. The Raman data are normalized to the Si substrate peak at 520 $cm^{-1}$ (not shown). Inset displays the full-width-half-maximum (FWHM) of the E' mode as a function of channel width. **b**, Averaged tip-enhanced photoluminescence (TEPL) spectra comparing a nanoribbon channel region to the much wider anchor region of the same nanoribbon, showing minimal broadening of the A-exciton peak and a slight increase in the trion-to-exciton intensity ratio. **c**, TEPL intensity map of an entire ~75 nm wide and ~1 μm long nanoribbon, with uniform optical emission along the channel. The black spot in the upper third of the nanoribbon is a measurement artifact. **d**, Scanning transmission electron microscopy (STEM) image of two parallel monolayer $MoS_2$ nanoribbons (from an array transferred onto $SiN_x$ membrane). Insets show diffraction patterns of three regions, revealing good crystallinity in the channel and near the edge. Inset scale bar 5 $nm^{-1}$. **e**, Magnified edge region in TEM mode from the same STEM.

## High-κ dielectric integration

To reduce the operating voltage of our nanoribbon transistors, we integrate ultrathin high-κ gate dielectrics into the gate stack. Achieving this involves transferring monolayer films onto pre-patterned local back-gates with $HfO_2$ dielectric (~1.5 nm equivalent oxide thickness), then applying our optimized nanoribbon process described earlier. (See **Methods** for additional details.) The schematic of the resulting nanoribbon devices is shown in **Fig. 4a** and confirmed by top-down SEM in **Fig. 4b**.

Here, we expand beyond $MoS_2$ by also examining monolayer *n*-type $WS_2$ and *p*-type $WSe_2$ nanoribbons, all patterned down to ~50 nm widths. Raman spectroscopy (**Supplementary Fig. 10**) indicates that the nanoribbon fabrication steps do not appear to introduce additional damage. **Figure 4c** compares the transfer characteristics of three such nanoribbon channels that are just 50 nm × 50 nm. Among these, monolayer $MoS_2$ achieves the highest on-state current density, $I_{max} \approx 460$ μA μm$^{-1}$ at $V_{DS} = 1$ V, albeit with negative $V_T$ (*i.e.,* normally-on, depletion-mode device). A ~60 nm wide monolayer $MoS_2$ nanoribbon reaches $I_{max} \approx 560$ μA μm$^{-1}$ at $V_{DS} = 1$ V, as shown in **Supplementary Fig. 11**.

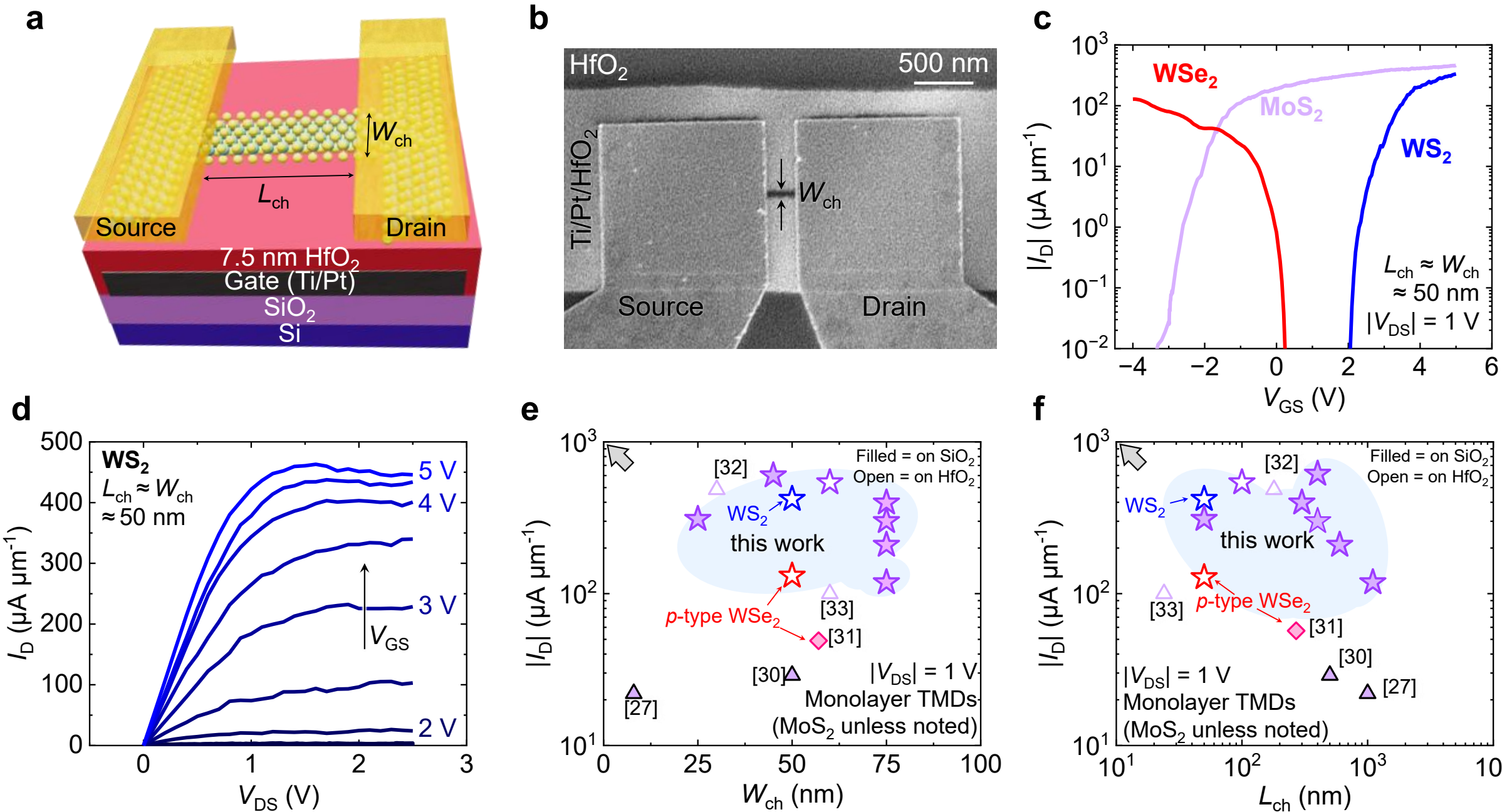

**Figure 4 | Complementary monolayer TMD nanoribbons with high-κ dielectric. a**, Schematic of monolayer nanoribbon including anchored contacts, here with thin $HfO_2$ dielectric. Figure not to scale. **b**, Top-down SEM image of a representative nanoribbon transistor. **c,** Measured transfer characteristics of monolayer $MoS_2$, $WS_2$, and $WSe_2$ nanoribbons with high-κ dielectric, all with channel length and width of ~50 nm. **d**, Measured output characteristics of monolayer $WS_2$ nanoribbon, showing well-behaved, normally-off (enhancement-mode) operation with high current saturation. **e**, Comparing $|I_{max}|$ of single-gated monolayer TMD nanoribbons vs. channel width, at $|V_{DS}| = 1$ V and maximum $|V_{GS}|$. Unlabeled symbols are $MoS_2$, while $WS_2$ and $WSe_2$ are labeled. Our devices, marked by star symbols, reach some of the highest current densities reported to date. Our $WS_2$ (blue star) has the highest $I_{max} \approx 420$ μA μm$^{-1}$ to date (at $V_{DS} = 1$ V) in a monolayer nanoribbon of this 2D semiconductor. Filled markers are on $SiO_2$ back-gate substrates[27–31]; open markers are devices with high-κ gate dielectric[32,33]. Symbols with red border are *p*-type $WSe_2$, all others are *n*-type. **f**, Additional benchmarking of single-gated nanoribbon devices, here vs. channel length, $L_{ch}$. Block arrows point to the desirable corner in both benchmarking plots.

On the other hand, monolayer $WS_2$ nanoribbons show a desirable positive $V_T$ (*i.e.,* normally-off, enhancement-mode device) but still reach $I_{max} \approx 420$ μA μm$^{-1}$ at $V_{DS} = 1$ V (and up to ~460 μA μm$^{-1}$ at $V_{DS} = 1.5$ V in **Fig. 4d**). These $WS_2$ current densities are the highest to date, by > 100×, for a nanoribbon of this material, likely due to our fabrication process and our use of stressed Ni/Au contacts[34] with good $R_c \approx 675 \pm 268$ Ω·μm (see **Supplementary Fig. 12**). Finally, the *p*-type $WSe_2$ nanoribbons reach $|I_{max}| \approx 130$ μA μm$^{-1}$, an encouraging result given its desirable negative $V_T$ (*i.e.,* normally-off, enhancement-mode device) and the historical performance gap between *p*-type and *n*-type TMD transistors. **Figure 4d** shows the output curves of the 50 nm × 50 nm $WS_2$ nanoribbon, with good current saturation and device turn-off at zero gate voltage, essential for future circuit implementation.

**Figures 4e,f** compare our results with other single-gate monolayer TMD nanoribbons to date, at $|V_{DS}| = 1$ V. Here, we benchmark current density, $|I_{max}|$, rather than mobility or contact resistance, because $I_{max}$ is less prone to measurement error (its greatest uncertainty comes from the nanoribbon width) and because, in principle, the threshold voltage can be adjusted by gate stack engineering[35,36]. $I_{max}$ also incorporates information about contact resistance and mobility, and it is ultimately responsible for circuit delays, which are inversely proportional to current density[37]. Our monolayer $MoS_2$ nanoribbons match or exceed previously-reported $I_{max}$ at comparable channel widths, while our monolayer $WS_2$ and $WSe_2$ nanoribbons exceed existing results, the $WS_2$ by a factor of > 100×. (We provide additional output curves, current drive statistics, and $V_T$ discussion across 140 devices in **Extended Data Figs. 2-4**.) Although sub-50 nm and even sub-10 nm wide nanoribbons (monolayer and multilayer), have been demonstrated using bottom-up[27,38–41] or anisotropic etching techniques[42], such devices have not reached the current densities of our top-down patterned nanoribbons at comparable channel widths.

While we have shown that the on-state of these nanoribbons can be as good as existing micrometer-scale devices, a remaining aspect is to inquire whether the off-state is affected by their edges. To probe this regime, we measured arrays of $MoS_2$ nanoribbons, as shown in **Extended Data Fig. 5**. These reach $I_{max}/I_{min}$ current ratios $>10^9$, limited by the measurement noise floor, also comparable with some of the best-known micrometer-scale devices to date. In other words, we conclude that, at least down to the channel widths probed here, there is no measurable edge conduction in the off-state, likely due to the mixed character[16,43] of our edges (see **Fig. 3c**). To our knowledge, no previous efforts on parallel nanoribbon arrays have probed the deep off-state, albeit a previous study[44] on random networks composed of narrower (10 to 30 nm) ribbons also found no evidence of off-state degradation. We provide additional comparisons of monolayer TMD nanoribbons, including some with unconventional geometry or fabrication approaches, in **Supplementary Table 1**.

## Conclusions

We demonstrate both *n*- and *p*-type (*i.e.,* complementary) nanoribbon transistors with $MoS_2$, $WS_2$, and $WSe_2$ at the ultimate limit of monolayer channel thinness. These achieve record-high current densities in channels down to ~25 nm widths, with the $WS_2$ nanoribbons in particular showing desirable, normally-off (enhancement mode) behavior, with good current saturation (~460 μA $\mu m^{-1}$ at 1.5 V drain-to-source voltage). The nanoribbons were enabled by mechanically robust "anchored" contacts which improve yield, a low-dose multi-patterning strategy with low-residue resist, and minimal edge degradation, studied by advanced nanoscale imaging, including tip-enhanced photoluminescence. Our work shows that scaling monolayer TMD channels down to ~25 nm widths does not degrade their on- or off-state, and such nanoribbon demonstrations are more technologically-relevant than micrometer-wide devices. Looking ahead, the compatibility with high-κ dielectrics, complementary device polarity, and good performance across several TMDs position monolayer nanoribbons as important building blocks of future gate-all-around[9] nanosheet transistors.

**Acknowledgements**. T.P., H.S., L.H. and E.P. were supported by SUPREME, a JUMP 2.0 Centre sponsored by the Semiconductor Research Corporation (SRC) and DARPA. T.P. would like to thank the NSF MPS-Ascend Postdoctoral Fellowship. A.E.O.P. acknowledges the Knut and Alice Wallenberg Foundation (grant number 2022.0374). T.P. and E.P. acknowledge Intel Corporation. K.N. acknowledges the Stanford Graduate Fellowship. J.A.Y. is supported by the Achievement Rewards for College Scientists Foundation. K.N. and J.A.Y. both thank the National Science Foundation Graduate Research Fellowship Program (grant number DGE-1656518). A.T.H. and A.J.M. acknowledge support from TSMC through the SystemX Alliance. Fabrication and characterization in this work were primarily performed at nano@stanford (RRID: SCR_026695). E.P. and A.J.M. acknowledge the NSF Future of Semiconductors (FuSe2) Programs (Award number 2425218). T.P., A.E.O.P. and E.P. acknowledge partial support from Samsung Electronics. We greatly appreciate A. Barnum for guidance with TEM imaging and Z. Han for thermal simulation discussions.

**Author contributions**. T.P. and A.E.O.P. contributed equally; E.P. conceived the project together with T.P. and A.E.O.P.; T.P., A.E.O.P., and A.T.H. synthesized the monolayer $MoS_2$ and Z.Z. grew the monolayer $WS_2$ samples; T.P. fabricated the devices, with process development guidance from A.E.O.P., K.N., and E.P.; H.S., Y.-M.L., and J.A.Y. assisted with various fabrication steps; L.H. led chloroform doping for the monolayer $WSe_2$ devices, with supervision from A.J.M.; T.P. and A.E.O.P. performed all electrical characterization; T.P. acquired micro-Raman and photoluminescence data, the atomic force microscopy measurements, and scanning electron imaging; A.K. conducted all tip-enhanced photoluminescence experiments; Á.F. carried out all transmission electron microscopy experiments, with supervision from P.C.M.; Y.S.S. and H.S. performed thermal simulations, with supervision from S.X.W. and E.P.; T.P., A.E.O.P., and E.P. wrote the manuscript, with input from all authors.

**Competing interests.** The authors declare the following competing financial interest(s): HORIBA Scientific is the manufacturer of the optical spectroscopy equipment used in this study. Collaboration with industry and academia is a part of A.K.'s job responsibilities. The remaining authors declare no additional financial or non-financial conflicts of interest.

**Data availability.** All data needed to evaluate the conclusions in this paper are present in the main text or the supplementary materials.

## Methods

**TMD synthesis and transfer.** All monolayer $MoS_2$ and $WS_2$ films were grown onto thermal 96 nm $SiO_2$ / Si and sapphire substrates respectively, as described in previous works[45,46]. The chemical vapor deposition grown monolayer $WSe_2$ on sapphire was purchased from 2D Semiconductors. For devices on $HfO_2$ (either 5.5 nm or 7.5 nm thick) with local back-gates, all monolayer films were transferred from their growth substrates, by spinning polystyrene (PS) at 1250 rpm for 60 seconds onto the growth substrate, followed by baking at 85 °C for 5 minutes. The TMD/PS stack was then immersed in deionized water to delaminate it from the growth substrate. The TMD/PS stack was then placed onto the chip with local back-gate structures, dried with $N_2$, and left in a $N_2$ dry box overnight. The following morning, the chips were heated at 85 °C for one hour, 150 °C for one hour, then cooled to room temperature. The chips were then placed in toluene overnight, followed by acetone and isopropanol cleaning (10 minutes each) to remove the PS from the surface. Before device fabrication, the chips were annealed in vacuum ($\sim10^{-5}$ torr) at 200 °C for two hours to promote TMD adhesion to the $HfO_2$.

**Fabrication process.** For the devices in **Figs. 1-2** and **3a**, the monolayer $MoS_2$ films were grown by solid-source chemical vapor deposition directly onto dry thermal $SiO_2$ (96 nm) on $p^{++}$ Si (resistivity < 5 mΩ·cm) substrates and were processed with no transfer, thus having better adhesion and higher yield. For these devices, electron-beam (e-beam) lithography (Raith EBPG 5200+ with 100 kV accelerating voltage) was used to first define coarse probing pads, which consist of e-beam evaporated $SiO_2$ (20 nm) / Ti (1 nm) / Pt (15 nm), where the $SiO_2$ is used to limit probing pad leakage to the Si back-gate. To construct the local back-gate samples in **Fig. 4a-d**, back-gate metals of Ti (1 nm) / Pt (13 nm) are defined by lift-off, then 7.5 nm $HfO_2$ is deposited by atomic layer deposition at 200 °C.

All nanoribbon channels were then defined using e-beam lithography and etched with $XeF_2$. A high-resolution AR-P 6200.04 (CSAR 62) resist was spun (3000 rpm, 60 s) and baked at 120 °C (5 minutes), yielding a ~50 nm resist thickness. Compared to conventional polymethyl methacrylate (PMMA) recipes, we found that the CSAR 62 resist allows for lower writing doses (here we use 425-475 μC cm$^{-2}$) and leaves less residue on the TMD surface. The resist was developed in room-temperature xylenes (45 s), followed by a quick isopropanol dip. This procedure would be repeated if implementing the multi-patterning (or LELE) approach (**Extended Data Fig. 1**). We found that colder or more dilute developer can further improve resolution (even using the single patterning [LE] process), but at the cost of requiring higher doses that risk TMD damage[21,22]. After channel formation, we defined fine source/drain contacts by a third e-beam lithography step and deposited using e-beam evaporation (~$10^{-8}$ torr). For monolayer $MoS_2$, ~40 nm Au contacts are employed, without an adhesion layer[6]. For monolayer $WS_2$, stressed Ni (10 nm) / Au (20 nm) fine contacts were used, following our previous work[34], to obtain good $R_c$. For monolayer $WSe_2$, Pd (10 nm) / Au (20 nm) contacts are utilized to promote hole injection, given the higher work function of Pd. Monolayer $WSe_2$ devices in this work were immersed in chloroform overnight before measurements, as chloroform has been found to lower $R_c$ for holes[47] in monolayer $WSe_2$. **Extended Data Figs. 2-4** have all three monolayer TMDs with Au (35 nm) contacts and with a 5.5 nm $HfO_2$ back-gate dielectric, where both *n*-type semiconductors experience a vacuum anneal and the *p*-type semiconductor is immersed in chloroform overnight before measurements.

**Electrical measurements.** Unless otherwise stated, electrical measurements were performed at room temperature using a Janis ST-100 vacuum probe station at ~$10^{-4}$ torr, with a Keithley 4200A semiconductor parameter analyzer. All monolayer $MoS_2$ devices presented in this work were first annealed at 250 °C for 2 hours under vacuum inside the probe station, to improve $R_c$ and remove adsorbates from the channel[6]. The monolayer $MoS_2$ devices are then measured after cooling back to room temperature, without breaking vacuum. Monolayer $WS_2$ devices presented in main text **Fig. 4** did not undergo any annealing procedures (to retain the strain state from the Ni/Au contacts), while the $WS_2$ devices shown in **Extended Data Figs. 2** and **3** experienced an anneal to improve their Au contacts. All monolayer $WSe_2$ devices presented did not experience any annealing prior to electrical testing in vacuum. Cryogenic measurements were conducted using a Lakeshore cryoprobe station under vacuum (~$10^{-6}$ torr) and a Keithley 4200A semiconductor parameter analyzer.

**Material characterization.** Micro-Raman spectroscopy was performed using a Horiba LabRAM instrument with a 532 nm laser with 1800 spectrometer grating at a laser power of 120 μW. Measurements were performed at room temperature and in ambient conditions. Atomic force microscopy (AFM) was conducted using a Bruker Dimension Icon in standard tapping mode with a NSC18 Pt probe. Scanning electron microscopy (SEM) was performed using either a FEI Magellan or Helios, with an accelerating voltage of 2 kV and beam current of 43 pA.

After the nanoribbons were visualized by AFM and/or SEM, we extracted the average nanoribbon width across the channel. This channel extraction was systematically conducted using ImageJ analysis software, where an average contrast line profile across the substrate and nanoribbon can be determined. This contrast line profile can be fitted to a Gaussian function, where the nanoribbon width is defined as the full-width-half-maximum of the fitted function. The fitting allows us to simultaneously calculate the uncertainty in our channel width estimates, which was typically 3-5 nm.

High-resolution transmission electron microscope (TEM) images were taken at 80 kV using a Thermo Fisher Spectra 300 operated in monochromated TEM imaging mode. To account for sample drift during acquisition, 80 frames were collected with drift correction enabled. The frames were then aligned and averaged to compensate for drift and improve the signal-to-noise ratio. Imaging was performed using a collection angle of 228 mrad. Final images were processed using a Drift-Corrected Frame Integration routine on the Ceta camera, which registers and integrates the aligned frames. A radial Wiener filter was applied to further suppress high-frequency noise and enhance image clarity. Energy-dispersive X-ray spectroscopy was performed in scanning transmission electron microscopy mode at 80 kV on the same instrument. Electron energy loss spectroscopy measurements were performed in single-mode, high-quality acquisition using the EF-CCD detector. The microscope was operated with a 50 μm C2 aperture, corresponding to a convergence semi-angle of 21.4 mrad and a camera length of 29 mm. The beam current was set to 0.213 nA. A smaller entrance aperture was used to improve energy resolution. The zero-loss peak was aligned prior to acquisition, and the spectrometer was tuned for optimal focus. The final zero-loss peak full width at half maximum was 1.05 eV, confirming good energy resolution. The dispersion was set to 0.3 eV per channel, suitable for resolving the C–K, N–K, O–K, and F–K edges while maintaining sufficient signal intensity. The monolayer $MoS_2$ samples for this experiment (**Fig. 3d,e**) were grown directly onto 300 nm $SiO_2$ on Si substrates, then underwent e-beam lithography and dry etching (following the channel definition procedure described above) to produce nanoribbons. The finalized nanoribbons were delaminated with a droplet of deionized water onto a polydimethylsiloxane stamp. The nanoribbons were then dry transferred from the stamp down onto a 10 nm thick $SiN_x$ TEM window (Norcada TA301Z).

Tip-enhanced photoluminescence (TEPL) maps were collected on LabRAM-Nano AFM-Raman system (HORIBA Scientific) modified for the concurrent excitation and collection with two lasers simultaneously[48]. Excitation and collection of the Raman signal was done using the side 100x, 0.7 NA objective (Mitutoyo) inclined at 25 degrees to the plane of the sample. Laser power on the sample for both 633 nm and 594 nm excitations were about 150 μW. TEPL maps were collected using Omni-Access-NC-Au (APPNano) TERS probes in DualSpec$^{TM}$ version of the SpecTop mode where in each pixel of the map spectra were collected with the tip in direct contact with the sample (near + far field) and in tapping operation with amplitude of about 20 nm (far field). Far field data were subtracted from the combined map to produce the pure near-field response. The monolayer $MoS_2$ samples for this experiment (main text **Fig. 3b,c**) were grown directly onto 300 nm $SiO_2$ on Si substrates, then underwent e-beam lithography and dry etching (following the channel definition procedure described above) to complete the nanoribbons. The samples were vacuum annealed at ~$10^{-4}$ torr at 250 °C for 8 hours to remove adsorbates and resist residues prior to experiments.

## Methods & Extended Data references

## 1. Multi-patterning (litho-etch-litho-etch, LELE) procedure

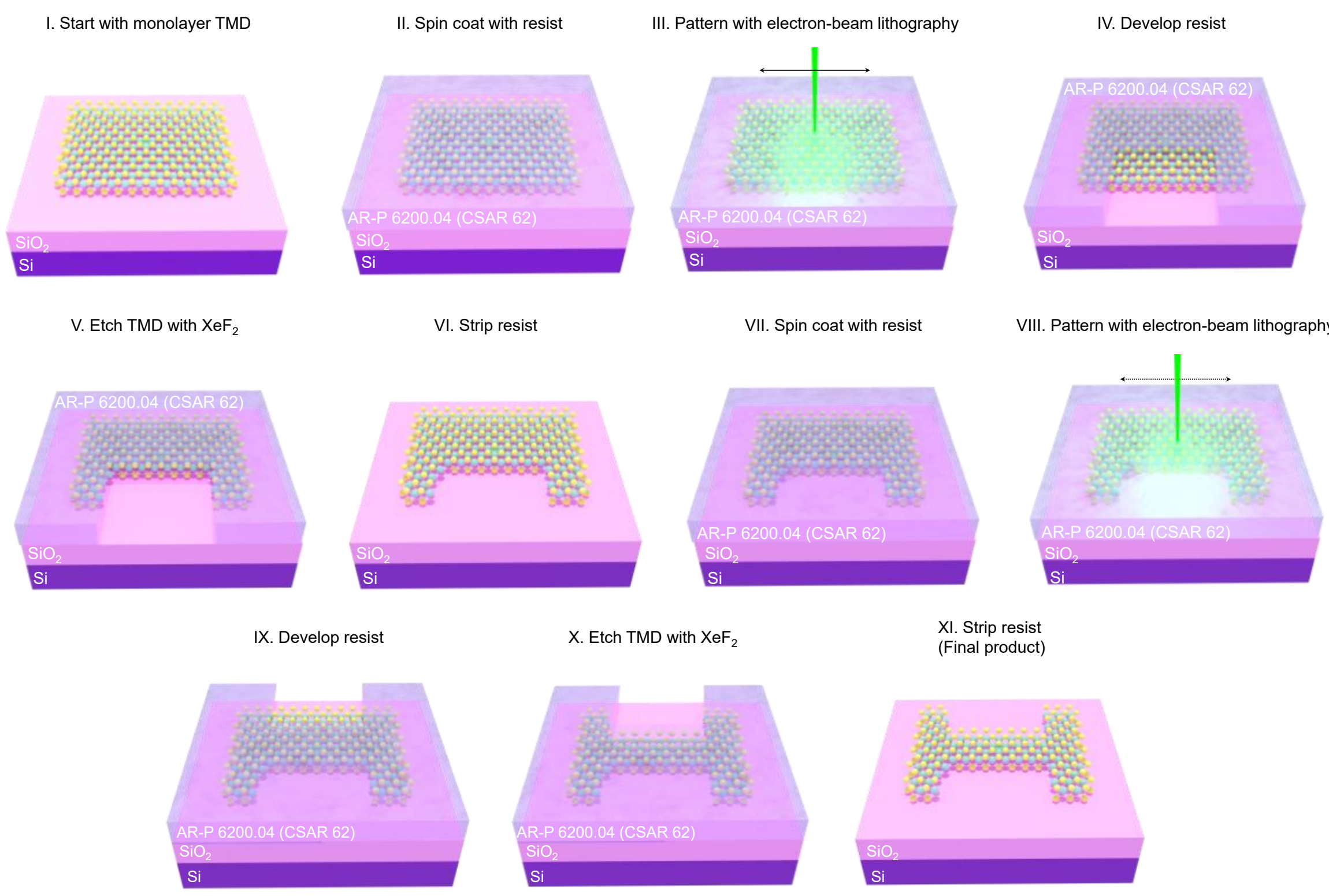

**Extended Data Figure 1 | Litho-etch-litho-etch (LELE) patterning of monolayer TMD nanoribbons.** The process involves two sequential e-beam lithography and etch steps. This dual-exposure approach enables sub-30 nm features by overcoming proximity-effect limitations inherent to single-step patterning, while maintaining the same (overall) low dose. See **Methods** for more information about doses and low-residue resist used for lithography.

## 2. Off-state current and output characteristics for various monolayer transition metal dichalcogenide (TMD) nanoribbons

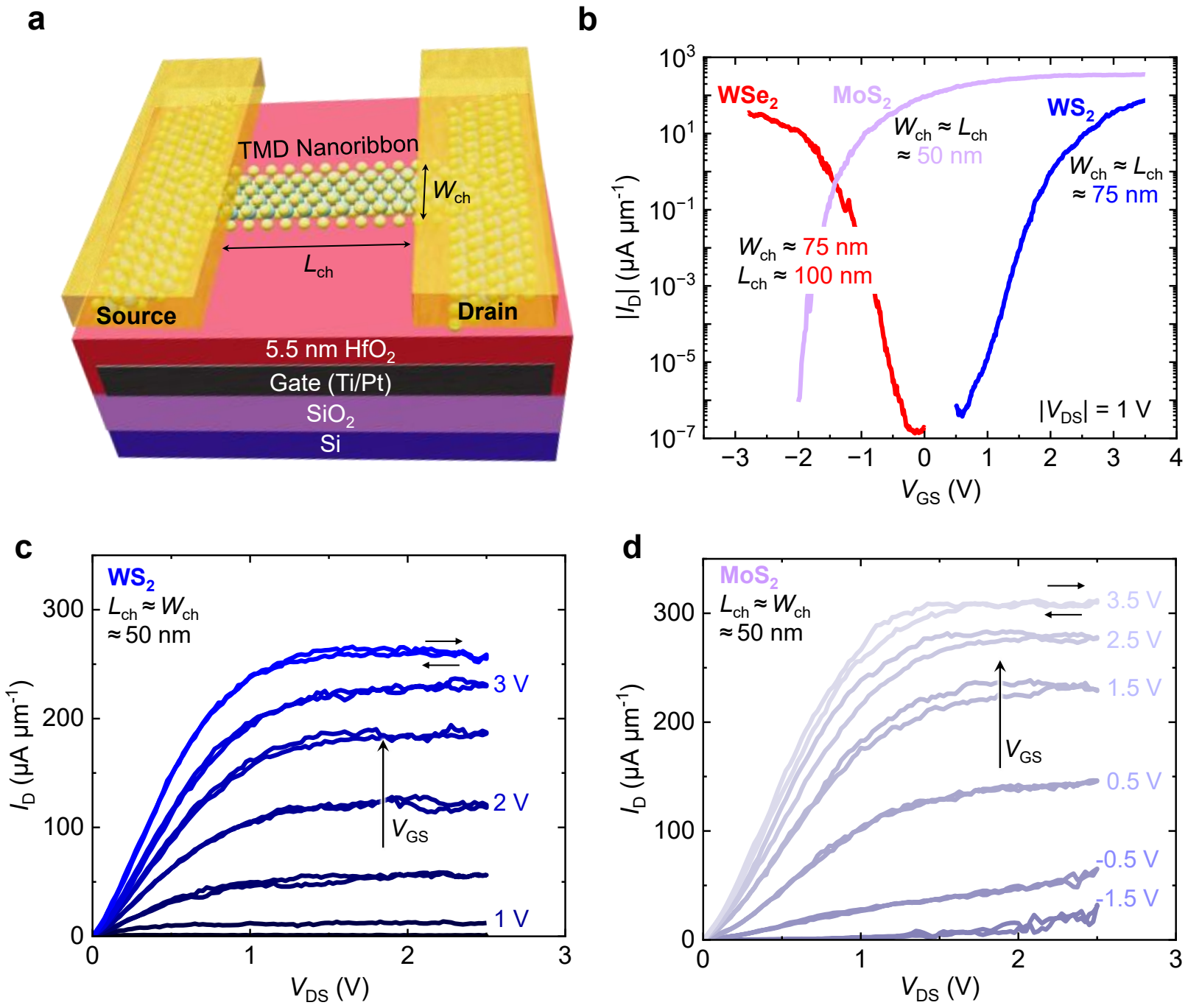

**Extended Data Figure 2 | Off-state current and output characteristics for various monolayer transition metal dichalcogenide (TMD) nanoribbons on 5.5 nm $HfO_2$ back-gate dielectrics. a,** Device schematic, for various TMD nanoribbons with Au contacts on 5.5 nm $HfO_2$ back-gate dielectric. Note that the $HfO_2$ dielectric on these test chips was slightly thinner than the 7.5 nm thick $HfO_2$ used in **Fig. 4** of the main text. **b,** Measured transfer characteristics of monolayer $MoS_2$, $WS_2$, and $WSe_2$ nanoribbons with Au source and drain contacts, on local back-gates with high-κ dielectric, including the deep off-state. Note that Au source-drain contacts are optimal for $MoS_2$ while Ni/Au and Pd/Au are preferable for $WS_2$ and $WSe_2$, respectively, explaining the lower on-currents seen here compared to **Fig. 4c** in the main text. Nevertheless, these each display $I_{max}/I_{min}$ ratios over $10^8$. **c,** Output characteristics of a $WS_2$ nanoribbon with dimensions and voltages as listed. **d,** Output characteristics of a $MoS_2$ nanoribbon with dimensions and voltages as listed. Both nanoribbons here had Au source and drain contacts, for direct comparison. (In contrast, $WS_2$ nanoribbons in main text **Fig. 4** had Ni/Au contacts.) Small arrows indicate voltage sweep directions, illustrating small hysteresis. All measurements are at room temperature.

## 3. Width-dependent current density across TMDs

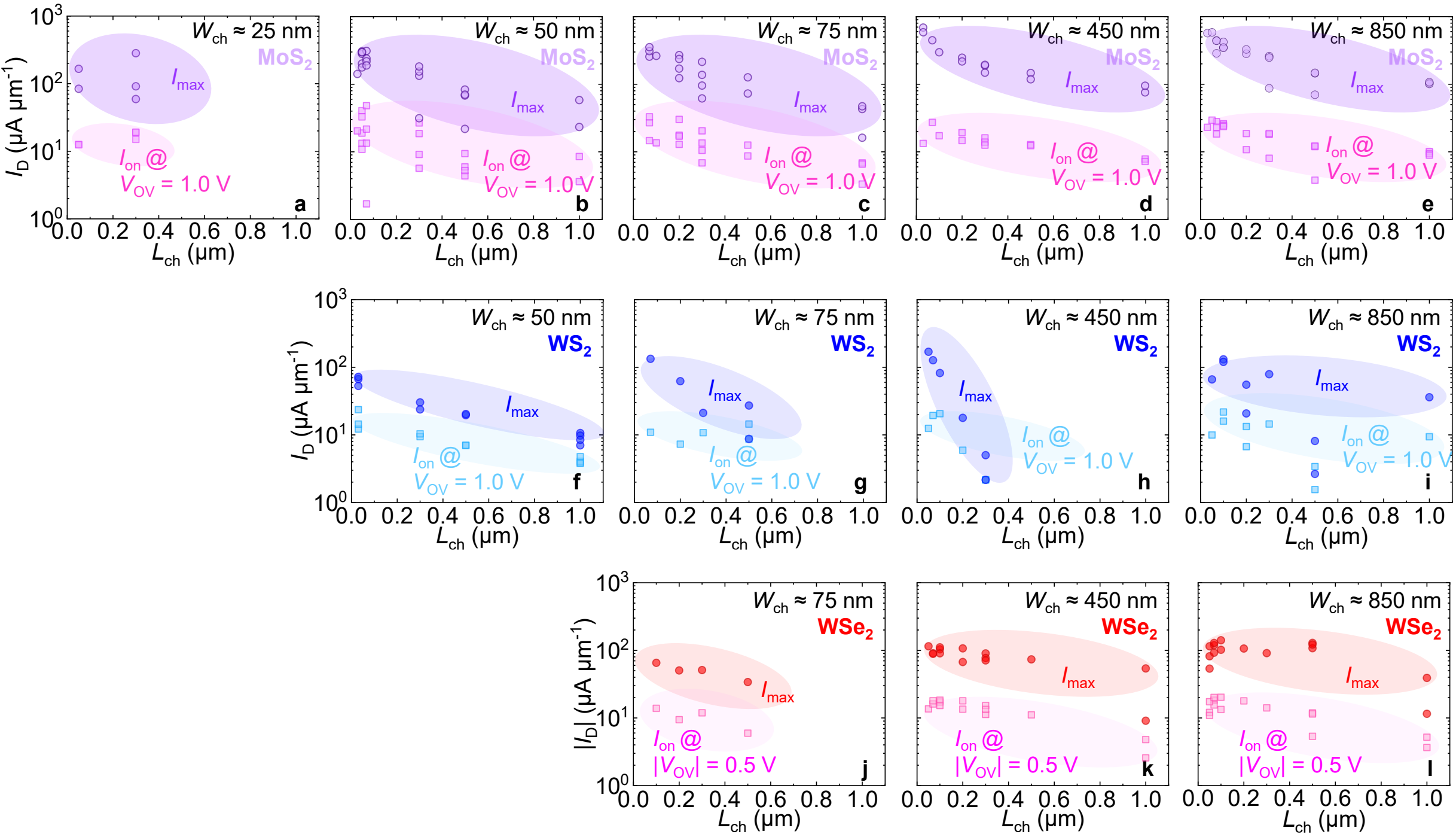

**Extended Data Figure 3 | Width-dependent current density across our monolayer TMDs.** Drain current density ($I_D$) vs. channel length ($L_{ch}$) for monolayer *n*-type $MoS_2$ devices with channel widths ($W_{ch}$) of approximately **a,** 25 nm, **b,** 50 nm, **c,** 75 nm, **d,** 450 nm, and **e,** 850 nm. Corresponding data for monolayer *n*-type $WS_2$ with $W_{ch}$ of approximately **f,** 50 nm, **g,** 75 nm, **h,** 450 nm, and **i,** 850 nm. Corresponding data for monolayer *p*-type $WSe_2$ with $W_{ch}$ of approximately **j,** 75 nm, **k,** 450 nm, and **l,** 850 nm. All devices in this figure have Au drain and source contacts, and were fabricated on 5.5 nm $HfO_2$ with local back-gates (see **Extended Data Fig. 2a**), measured at $|V_{DS}| = 1$ V. Each panel shows both $I_{max}$ (at maximum $|V_{GS}| = 3$ V for $WSe_2$ and 3.5 V for the others) and $I_{on}$ at $V_{ov} = |V_{GS} - V_{T,cc}| = 1$ and 0.5 V, for *n*-type and *p*-type films respectively, where $V_{T,cc}$ is taken at $I_D = 100$ nA/μm. Within the device-to-device variability typical of academic nanofabrication, we observe no obvious changes in current density between nanoribbons (left columns) and microribbons (right columns) of a given material, after accounting for $V_T$ changes. The slightly higher $I_{max}$ in wider channels of $MoS_2$ and $WS_2$ is due to their lower $V_{T,cc}$ (see **Extended Data Fig. 4** for more discussion); accordingly, no obvious nano- vs. microribbon difference is seen in $I_{on}$. Note, some nanoribbon widths are missing for $WS_2$ and $WSe_2$ due to adhesion and yield issues, from the layer transfer process onto $HfO_2$ with local back-gates.

## 4. Threshold voltage vs. channel length and width for monolayer $MoS_2$ on $HfO_2$

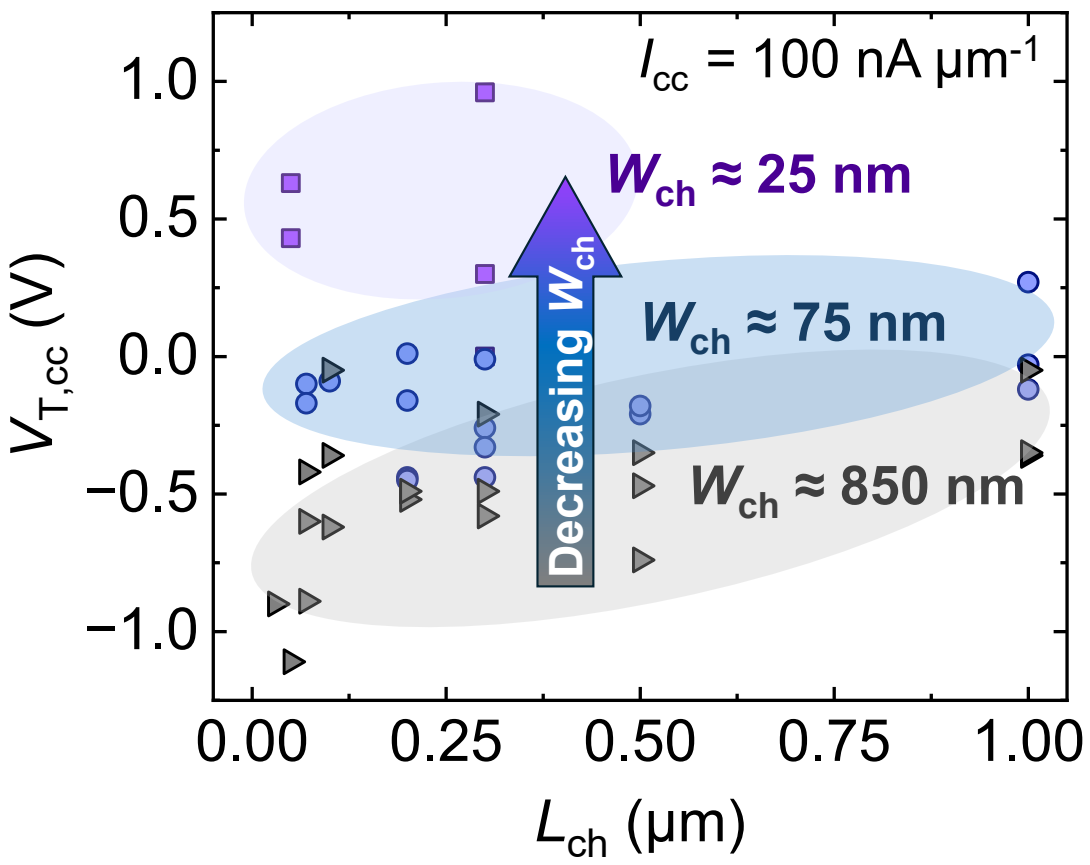

**Extended Data Figure 4 | Threshold voltage as a function of channel length and width**, for our monolayer $MoS_2$ devices on $HfO_2$ from **Extended Data Figs. 3a-e**. Displayed are devices with channel widths $W_{ch}$ ≈ 850 nm (grey triangles), 75 nm (blue circles), and 25 nm (purple squares). $V_{T,cc}$ is estimated at a constant current of 100 nA µm$^{-1}$. $V_{T,cc}$ increases as $W_{ch}$ is reduced, causing an apparent decrease of $I_{max}$ for the nanoribbons at maximum $V_{GS}$ (= 3.5 V here on ~5.5 nm $HfO_2$ back-gate dielectric). Our findings here are consistent with earlier work[49] on thick $MoS_2$ nanoribbons (~10 layers), which also found increased $V_T$ as $W_{ch}$ is reduced, attributing it to negative charges adsorbed at the edges. Fixed edge charges may arise from fabrication or atmospheric exposure, and are likely below EELS detection limits. On thick back-gate dielectrics (comparable to or thicker than $W_{ch}$), the relatively higher contribution of fringing fields at the edges could have a competing effect and *decrease* $V_T$ as $W_{ch}$ is reduced, as seen in **Supplementary Fig. 3c** for devices on 96 nm $SiO_2$. The analysis here was conducted for the forward-sweep of the transfer curves, however we note that the hysteresis was quite small for these devices (see **Extended Data Fig. 2d**), thus rendering the trends here to be robust to either sweep direction.

## 5. Nanoribbon array electrical performance

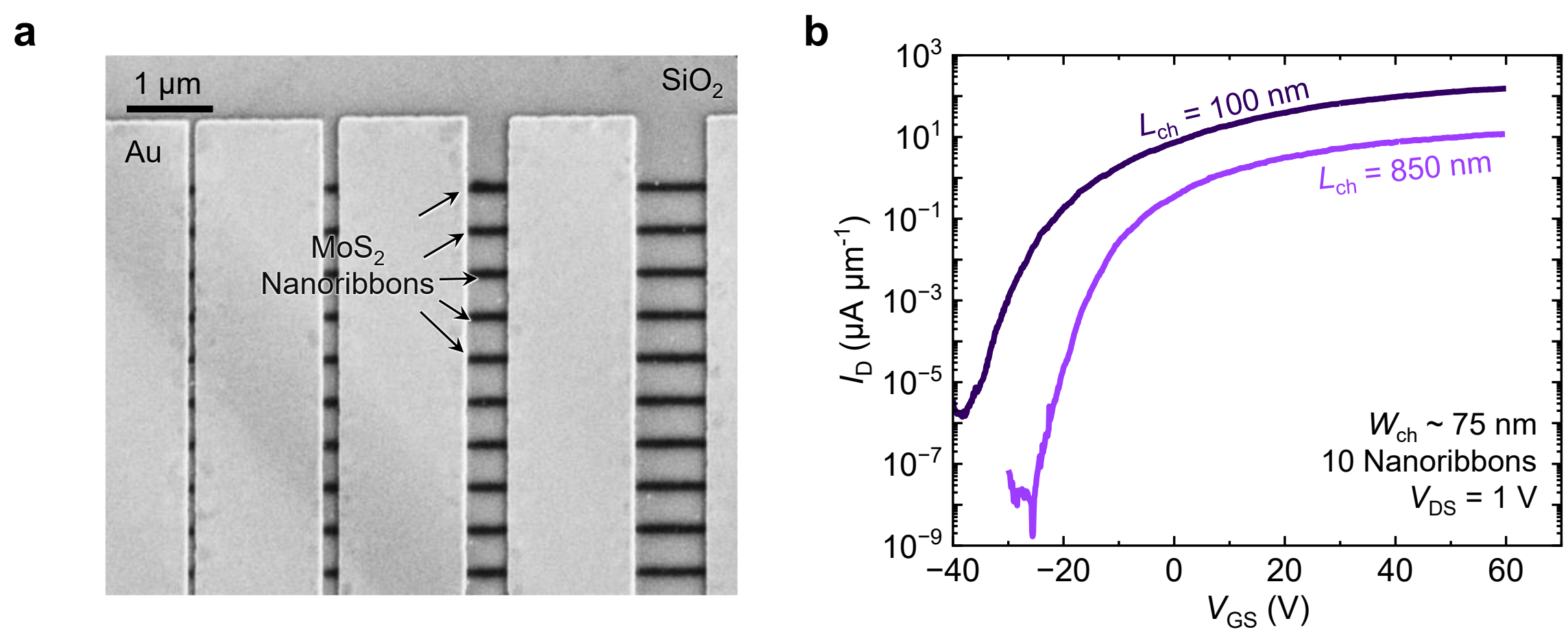

**Extended Data Figure 5 | Electrical performance of monolayer $MoS_2$ nanoribbon arrays. a**, Top-down scanning electron image of the nanoribbon array, showing widths down to ~75 nm and 10 parallel nanoribbons. The devices here are monolayer $MoS_2$ directly grown and fabricated onto 96 nm $SiO_2$ on $p^{++}$ Si. In the image, the nanoribbons appear dark, while the lighter large rectangles are the Au contacts. **b**, Measured transfer characteristics of short- and long-channel nanoribbon array devices. The on-state current density is comparable to that of single nanoribbons (see main text **Fig. 2a**). The off-state current density is one to two orders of magnitude lower than that measured in individual nanoribbons due to the width difference and differences in measurement settings, and comparable to some of the best-known micrometer-scale devices ($I_{max}/I_{min}$ current ratio > $10^9$). The short-channel device (100 nm here) exhibits slightly degraded off-state, likely from $V_T$ roll-off and other short-channel effects, on the thicker $SiO_2$ (96 nm) back-gate oxide.